# Forensic analysis of the Windows telemetry for diagnostics

Jaehyeok Han, Jungheum Park, Hyunji Chung and Sangjin Lee

Center for Information Security Technologies, Korea University
(145, Anam-ro, Seongbuk-gu, Seoul, South Korea)

**Abstract**

Telemetry is the automated sensing and collection of data from a remote device. It is often used to provide better services for users. Microsoft uses telemetry to periodically collect information about Windows systems and to help improve user experience and fix potential issues. Windows telemetry service functions by creating RBS files on the local system to reliably transfer and manage the telemetry data, and these files can provide useful information in a digital forensic investigation. Combined with the information derived from traditional Windows forensics, investigators can have greater confidence in the evidence derived from various artifacts. It is possible to acquire information that can be confirmed only for live systems, such as the computer hardware serial number, the connection records for external storage devices, and traces of executed processes. This information is included in the RBS files that are created for use in Windows telemetry. In this paper, we introduced how to acquire RBS files telemetry and analyzed the data structure of these RBS files, which are able to determine the types of information that Windows OS have been collected. We also discussed the reliability and the novelty by comparing the conventional artifacts with the RBS files, which could be useful in digital forensic investigation.

*Keywords:* Windows 10 forensics; Windows telemetry; diagnostic data; RBS file;

## 1. Introduction

Telemetry is used to periodically and remotely collect environmental data from a target system in order to provide better services. Its main functions are the automated sensing, measurement of data, and control of remote devices. It is commonly used in computers, automobiles, medical devices, and many industries [1]. A widespread example of this technology in use involves Microsoft, who first introduced telemetry to their ecosystem with the launch of Windows 10 under the service name 'DiagTrack'. This service is configured to send back various types of information, including the user experience and telemetry components, as a way of reporting errors, thus helping improve the quality of its services. Users can opt out of this service, but it is set to operate by default [2]. Since the usage share of Windows 7 and later versions accounts for more than 95% [3], most operating systems can be used functionally through the update.

Forensic analysis of Window telemetry is important for two reasons. First, it can provide the same results as those gained from other artifacts. Windows artifacts include the Master File Table (MFT), Registry, Web History, Event Log, Prefetch, shortcuts, and so on. By analyzing these artifacts in a digital forensic investigation, a collection of events organized as a timeline is able to be constructed. The RBS files used for Windows telemetry contain some of the same information as in others. This means that any examination of a target system will be more reliable than if the information was derived from only a single artifact. It also can be a potential countermeasure against anti-forensic techniques, such as data manipulation or erasure of the digital evidence. The second reason is that it can provide new information that cannot be extracted in a conventional investigation. For example, the unique value of hardware device identifiers, such as an HDD, an SSD, external storage or a RAM card, and running process lists are only visible during live forensics. This information can only be extracted from RBS files and it could be useful in a digital forensic investigation. In particular, Windows telemetry has not been studied so far.

Therefore, in this paper, we describe the data structure of these RBS files and their content in comparison to other Windows artifacts. We also discuss how they can be used in digital forensic investigations.

## 2. Background

What is the Windows telemetry? Traditional models using ICMP, SNMP have been used to collect operational statistics from a system or network, but have several restrictions due to the growth of the objects like IoT devices and cloud-based applications [4]. Responding to this limitation, telemetry is a new approach for monitoring and control where data is streamed continuously and provides almost real-time access to operational statistics. Automated sensing and measurement of data is sent from a remote device to a central control point, and in this case of the Windows telemetry it is sent to Microsoft.

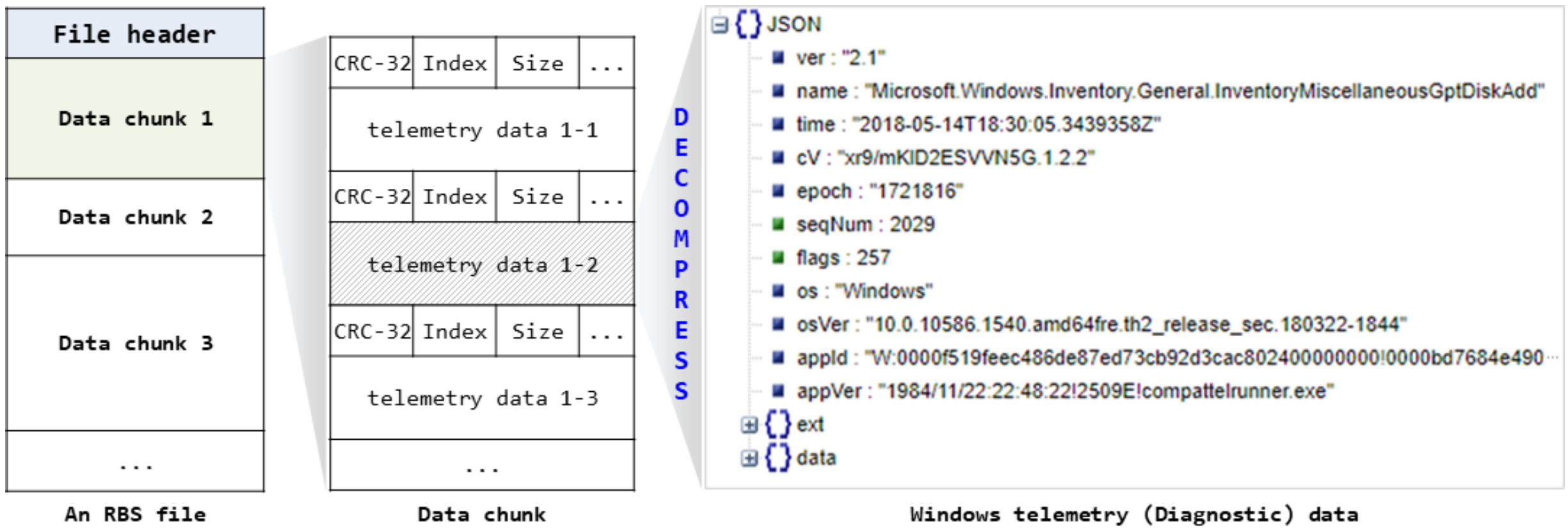

Figure 1. The overall structure of the Windows telemetry (RBS file) for the diagnostics

Windows telemetry with DiagTrack service works by calling ‘svchost.exe’ process with the file ‘%SystemRoot%\System32\diagtrack.dll’. Usually, it transmits telemetry data (i.e., diagnostic data) every 15 minutes, though this can vary depending on the power options and network settings. If a certain event occurs, DiagTrack checks its own status and writes the telemetry data to files with the .rbs extension in ‘%ProgramData%\Microsoft\Diagnosis\’ directory on the local system. Parts of these files are later sent as payloads with the HTTPS protocol based on TLS. This means that if an error occurs during system operation, the relevant data is stored in files first and then sent at the scheduled time. Windows telemetry is categorized into four levels: Security, Basic, Enhanced, and Full [5]. The higher the level, the more information that is recorded in the local system and could be sent to Microsoft.

Similar to Windows telemetry, there is an office telemetry, which is a part of compatibility monitoring framework [6]. This data is collected by agent from office applications and it includes document’s details like a file name, file size, author, last updated time, and so on [7]. Since office telemetry data substantially enriches metadata it helps to create timelines [8], this will be useful in case of investigation for the enterprise controlled by organization. These office telemetry is stored in SQL server, however, Windows telemetry save in a new file format, which has an file extension of 'RBS', rather than to the SQL databases. Therefore, this paper describes the structure of the RBS file that stores Window telemetry and how to use digital forensic investigation.

## 3. Analysis of the RBS files

### 3.1 Acquisition and Format Internals

To acquire the RBS files from active (live) system, the related services should be terminated first and then copy the target. If there is a disk image file, the RBS files need to be identified by filesystem analysis and then extract them. RBS files have different names and a fixed size depending on the version of Windows being used, as shown in Table 1. The data structure and telemetry event data also differ between versions. Windows telemetry has been in use since the launch of Windows 10, but Windows 7 or later is also supported through an update. In this case, the data structure and characteristics are the same as those of the 1507 version. The file names and the fixed size have changed since Windows 10 version 1607. Depending on when you update Windows, both versions of files can exist. For example, if it was installed and had used the 1507 version and then updated to the 1803 version, then both 'events00.rbs' and 'Events_Normal.rbs' will be there.

RBS files are composed of a file header and data chunks, and different forms of telemetry data are recorded within a data chunk. Fig. 1 shows the structure of the typical RBS file. The number of data chunks is related to how long the system has been used. Because the size of the files is fixed, unused areas where telemetry data has never been recorded is padded with 0x00. If the telemetry data to be recorded exceeds the size of the corresponding file, it is overwritten from an offset after the file header where data from the past is stored. When the Windows telemetry service is terminated, the RBS files are retained, but no additional telemetry data is created. Upon restarting the service, new telemetry data is appended after the existing data.

Table 1 File names and sizes(kilobytes) of the RBS files

| File name | File type | Version of Windows 10 | | | | |
|---|---|---|---|---|---|---|
| | | 1507 | 1511 | 1607 | 1703 1709 | 1803 1809 |
| events00.rbs | 0x00 | 24,576 | 49,152 | - | - | - |
| events01.rbs | 0x01 | 6,226 | 12,452 | - | - | - |
| events10.rbs | 0x03 | 492 | 984 | - | - | - |
| events11.rbs | 0x02 | 1,475 | 2,950 | - | - | - |
| Events_Normal.rbs | 0x00 | - | - | 1,6384 | 1,6384 | 1,6384 |
| Events_NormalCritical.rbs | 0x01 | - | - | 6,554 | 6,554 | 6,554 |
| Events_CostDeferred.rbs | 0x02 | - | - | 6,554 | 6,554 | 6,554 |
| Events_Realtime.rbs | 0x03 | - | - | 3,277 | 3,277 | 3,277 |

Table 2. Data structure of the file header

| Offset | Size | Description |
|---|---|---|
| 0x00 | 8 | Signature<br>· 1507, 1511 versions: UTCRBES3<br>· 1607 version: UTCRBES5<br>· 1703 version: UTCRBES7<br>· 1709, 1803 versions : UTCRBES8 |
| 0x08 | 8 | Timestamp: Last modified time<br>· Windows: 64bit hex value – little endian |
| 0x10 | 4 | Start offset of the last written data chunk 1 |
| 0x14 | 4 | Start offset of the last written data chunk 2 |
| 0x18 | 4 | Size of the last written data chunk |
| 0x1C | 4 | Total size of the data chunks |
| 0x20 | 4 | Index of the last written data chunk 1 |
| 0x24 | 4 | Index of the last written data chunk 2 |
| 0x28 | 2 | file type<br>· 0x00: Events_Normal.rbs<br>· 0x01: Events_NormalCritical.rbs<br>· 0x02: Events_CostDeferred.rbs<br>· 0x03: Events_Realtime.rbs |
| 0x2A | 11 | reserved (for the 1607 version or higher) |

The file header (Table 2) is located at the beginning of the RBS file. The structure of the file header differs slightly between Windows 10 versions but it can be identified by a signature, which is 8 bytes and contains 'UTCRBES' in ASCII code. It has a timestamp which represents the last time the telemetry data was modified and contains an offset and index for storing new telemetry data. The offset and the index have two fields each. Most of them in the dataset have the same values; only when the service had halted while the samples were being collected, there was a slight difference. Contents in the telemetry data have the own attribution depending on the 'name' value, like all Windows Event Log have a unique event ID, and each RBS files have the certain contents by 'file type' values.

After the file header, the data chunks are stored, which is the area corresponding to the payload for the telemetry service. The telemetry data which generated by periodically monitoring the state of the system is compressed and then stored. The size of the generated data chunk is variable because the frequency of checking the system status varies according to the value of 'name'. Details of the data chunks and their fields are shown in Table 3.

Table 3. Data structure of the data chunk

| Offset | Size | Description |
|---|---|---|
| 0x00 | 4 | CRC-32 value of telemetry data |
| 0x04 | 4 | Index of the telemetry data |
| 0x08 | 4 | Size of the Base64 encoded data |
| 0x0C | 4 | Size of the DEFLATE compressed data |
| 0x10 | 4 | Count of the telemetry data |
| 0x14 | 2 | reserved |
| 0x16 | variable | Base64 encoded data |
| - | variable | DEFLATE compressed data (JSON) |

Base64 encoded data from the offset 0x16 uses the '=' padding character and it needs further analysis in the future because the decoded data with Base64 was highly random. For the compressed data in the data chunk, a string in the JSON [9] object format can be extracted by decompressing with DEFLATE algorithm [10], and this JSON string is consist of the collection of key/value pairs with several keys (ver, name, time, flag, os, osVer, appId, appVer, etc.) and two objects (ext and data). The telemetry data is contained in the content of the JSON object and is configured differently within this object according to the 'name' value. This can provide the basis for digital forensic analysis.

### 3.2 Attributions of the Windows telemetry

RBS files contains that can help to determine a user's behavior and their system's environmental settings in a digital forensic investigation. Furthermore, since some of them cannot be identified in a conventional investigation, we can call the rbs file a new artifact. Diagnostic Data Viewer [11], which has been supported since Windows 10 version 1803, can be employed to check the Windows telemetry, but this program is functionally limited because this works only for the active system and it shows the parts of stored data. In this section, we demonstrate how to obtain telemetry data from the RBS files.

For analysis and verification, we acquired 340 RBS files from Windows 7/8.1/10, which support the telemetry service, and identified a number of 'name' values, exactly 237 events, from more than two million telemetry data sets. The oldest timestamp was recorded a year earlier and we checked cases that telemetry data was overwritten if it's generated more than the fixed size of the RBS file. Diagnostic events and fields can be partially determined because Microsoft has released information about telemetry data [12-14], but we confirmed that is a part of the document and have studied to be used as new information in the digital forensic investigations. We uploaded the dataset of the telemetry data and the RBS file parser developed in this study. [https://github.com/JaehyeokHan/Windows-Telemetry]

```
C:\WINDOWS\system32\cmd.exe
C:₩>wmic diskdrive get DeviceID, SerialNumber, Index, Size, BytesPerSector, Model
BytesPerSector  DeviceID            Index  Model                     SerialNumber      Size
512             ₩₩.₩PHYSICALDRIVE1  1      ST4000DM004-2CV104        ZFN1S9C4          4000784417280
512             ₩₩.₩PHYSICALDRIVE2  2      Samsung SSD 850 PRO 256GB S1SUNSAFC82202K   256052966400
512             ₩₩.₩PHYSICALDRIVE0  0      Samsung SSD 860 EVO 1TB   S3YBNB0K100915R   1000202273280
```

Figure 2. An example of the hardware device identifier by WMIC command on the system in action (live). The red boxed one is the device identified in the telemetry of Figure 3.

**Property 1. OS and application life cycle.** Telemetry data contains information on Windows OS installation and the addition and removal of applications. Typically, the source of this information is the Registry [15] and the Event Log. The telemetry data contains information such as the product name, owner, Windows version, product ID, installation date, product keys for activation, which are same with the Registry's. The advantage here is that an investigator is able to gain a record of the point at which the data was acquired and thus gains an overview of the history of the target system. If Windows 7 updated to Windows 10, the Registry would change the installation time to the time of the update, rather than recording the time when Windows 7 was originally installed. In this case, the filesystem log(metadata) needs to be investigated, but telemetry data provides this history from the initial system because telemetry data records all life cycle of both installations and removals for the OS and applications.

**Property 2. Hardware device information.** Hardware device identifiers and connection traces are contained. In terms of hardware device identifiers, an information from the Registry has a limitation for the mainboard (base board), monitor, optical disc drive (ODD), and a part of the network interface controller (NIC). But telemetry data includes more detailed specification information, such as the model name, serial number, disk partitions state, mainboard, CPU, RAM, and all storage devices including internals and externals. First of all, this information about devices identifier is the same as the output of the Windows Management Instrumentation Command-line (WMIC) [16]. Fig. 2 and 3 are the example showing that the information of the telemetry data is same as the output of WMIC for the property 2. Also if the type and brand of digital devices needs to be documented [17], the telemetry data make it possible to confirm with what are actually checked.

User data from applications like Messenger is often encrypted and most of the encryption key is generated using a hardware-inherent value. In this case, hardware-specific information is able to give us chances to look into the decrypted user data. Additionally, it is useful to track devices connection history like the changes and the usage of external

```
"DeviceId": "\\\\.\\PHYSICALDRIVE0",
"SerialNumber": "S3YBNB0K100915R",
"Index": 0,
"Size": "1000202273280",
"NumPartitions": 3,
"BytesPerSector": 512,
"MediaType": "SSD",
```

Figure 3. Hardware device identifier from the telemetry data, which name is 'Microsoft.Windows.Inventory.General.InventoryMiscellaneousPhysicalDiskInfoAdd'

storage with a target system, in particular detail events about the driver installed (initial connection), connection and disconnection. Some of traces includes the Registry and Event Log [19].

**Property 3. Process execution history.** Process name and the timestamp of the running process is collected at the Enhanced and Full level. This property was only able to be identified with live forensics but the analysis of telemetry data makes it possible to collect process execution history. Previously, only the name of an executable, a count of how many times it has been run, and a timestamp indicating the last time it was run were available from Prefetch, Icon Cache [18], and Event Log. In case of the Event Log, security policy setting is particularly required to be established for a forensic readiness rather than leave the default. As a result of analysis using our dataset in the several system environment, this property were stored for about three month. Hence, a long-term patterns of the cyberattack such as APT (Advanced Persistent Threat) can remain and the execution of malicious code or portable programs would be detected. There were events with different 'name' values for applications developed by Microsoft such as Skype, Outlook, OneDrive, to diagnose them in shorter-term cycle compare to the third party application.

**Property 4. Boot sector and partition table.** The RBS files store boot timestamps and raw data of the boot sector data—MBR (Master Boot Record), VBR (Volume Boot Record) or GPT (GUID Partition Table). The MBR has a partition table so it is able to record changes in the partitions on a disk. When a system damaged by a boot sector virus [20], it can be restored using the boot record data stored as telemetry data.

Table 4. Classification of the properties by 'name' values and sources for the Windows telemetry formatted the JSON object

| Property | Name | | Source (file type) |
|---|---|---|---|
| OS and application life cycle | • Census.OS | | Events_Realtime.rbs (0x03) |
| | • Microsoft.Windows.Inventory.Core.InventoryApplicationAdd<br>• Microsoft.Windows.Inventory.Core.InventoryApplicationRemove<br>• Microsoft.Windows.Kernel.Power.OSStateChange | | Events_NormalCritical.rbs (0x01) |
| Hardware device information | • Census.Hardware<br>• Census.Storage | • Census.Memory<br>• Census.Processor | Events_Realtime.rbs (0x03) |
| | • Microsoft.Windows.Inventory.General.InventoryMiscellaneousPhysicalDiskInfoAdd<br>• Microsoft.Windows.Inventory.Core.InventoryDevicePnpAdd<br>• Microsoft.Windows.Inventory.Core.InventoryDevicePnpRemove<br>• Microsoft.Windows.Inventory.Core.InventoryDeviceContainerAdd<br>• Microsoft.Windows.Inventory.Core.InventoryDeviceContainerRemove<br>• Microsoft.Windows.Inventory.Core.InventoryDriverPackageAdd | | Events_NormalCritical.rbs (0x01) |
| Process execution history | • Win32kTraceLogging.AppInteractivitySummary<br>• Win32kTraceLogging.PostUpdateUseInfo<br>• Microsoft.Windows.Narrator.Asimov.NarratorCommandGeneratedStart<br>• Microsoft.Windows.HangReporting.AppHangEvent<br>• Microsoft.OneDrive.Sync.Updater.ComponentInstallState | | Events_NormalCritical.rbs (0x01) |
| | • Microsoft.Windows.Skype.Host.UserLoggedIn<br>• Microsoft-Windows-Store.StoreLaunching | | Events_Normal.rbs (0x00) |
| Boot sector and partition table | • Microsoft.Windows.Inventory.General.InventoryMiscellaneousMbrDiskAdd<br>• Microsoft.Windows.Inventory.General.InventoryMiscellaneousGptDiskAdd<br>• Microsoft.Windows.Kernel.Power.OSStateChange | | Events_NormalCritical.rbs (0x01) |

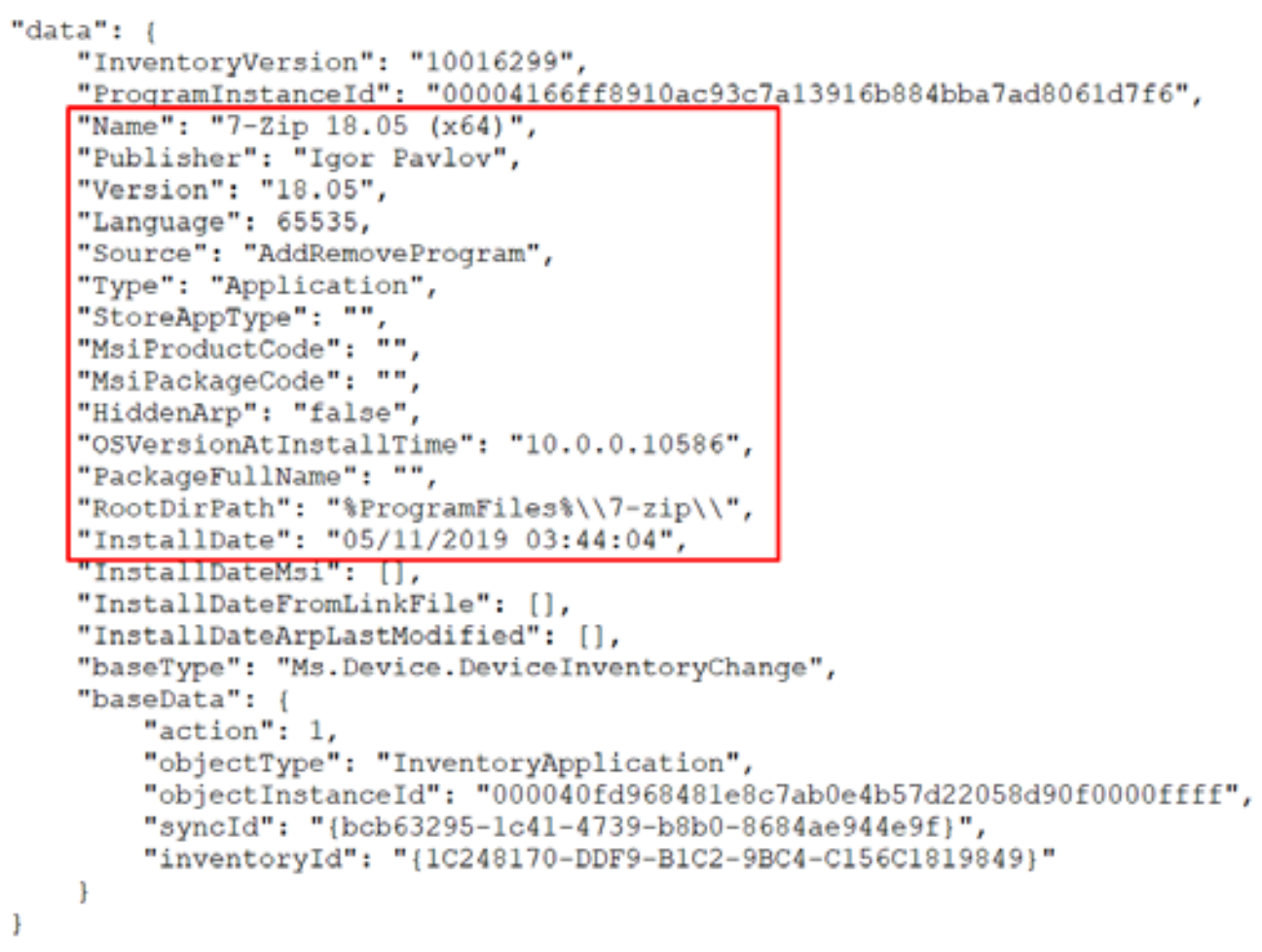

"data": {
    "InventoryVersion": "10016299",
    "ProgramInstanceId": "00004166ff8910ac93c7a13916b884bba7ad8061d7f6",
    "Name": "7-Zip 18.05 (x64)",
    "Publisher": "Igor Pavlov",
    "Version": "18.05",
    "Language": 65535,
    "Source": "AddRemoveProgram",
    "Type": "Application",
    "StoreAppType": "",
    "MsiProductCode": "",
    "MsiPackageCode": "",
    "HiddenArp": "false",
    "OSVersionAtInstallTime": "10.0.0.10586",
    "PackageFullName": "",
    "RootDirPath": "%ProgramFiles%\\7-zip\\",
    "InstallDate": "05/11/2019 03:44:04",
    "InstallDateMsi": [],
    "InstallDateFromLinkFile": [],
    "InstallDateArpLastModified": [],
    "baseType": "Ms.Device.DeviceInventoryChange",
    "baseData": {
        "action": 1,
        "objectType": "InventoryApplication",
        "objectInstanceId": "000040fd968481e8c7ab0e4b57d22058d90f0000ffff",
        "syncId": "{bcb63295-1c41-4739-b8b0-8684ae944e9f}",
        "inventoryId": "{1C248170-DDF9-B1C2-9BC4-C156C1819849}"
    }
}

**name: Microsoft.Windows.Inventory.Core.InventoryApplicationAdd**

This event was made when the 7-zip had installed. An information of the application (publisher, version, language) and the application installation timestamp are verified. If application comes from the Microsoft store are more specifically.

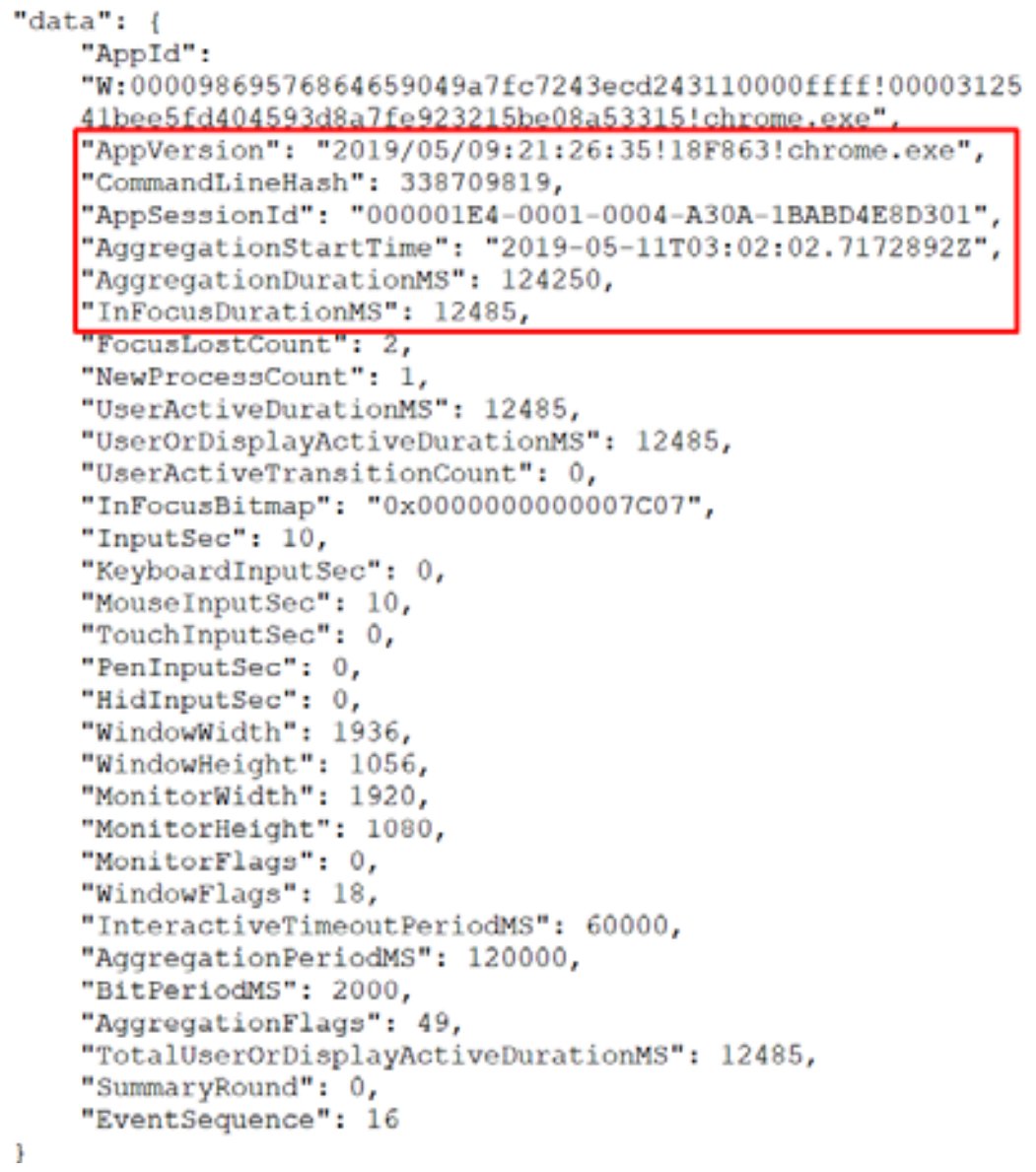

"data": {
    "AppId":
    "W:0000986957686465904 9a7fc7243ecd243110000ffff!00003125
    41bee5fd404593d8a7fe923215be08a53315!chrome.exe",
    "AppVersion": "2019/05/09:21:26:35!18F863!chrome.exe",
    "CommandLineHash": 338709819,
    "AppSessionId": "000001E4-0001-0004-A30A-1BABD4E8D301",
    "AggregationStartTime": "2019-05-11T03:02:02.7172892Z",
    "AggregationDurationMS": 124250,
    "InFocusDurationMS": 12485,
    "FocusLostCount": 2,
    "NewProcessCount": 1,
    "UserActiveDurationMS": 12485,
    "UserOrDisplayActiveDurationMS": 12485,
    "UserActiveTransitionCount": 0,
    "InFocusBitmap": "0x0000000000007C07",
    "InputSec": 10,
    "KeyboardInputSec": 0,
    "MouseInputSec": 10,
    "TouchInputSec": 0,
    "PenInputSec": 0,
    "HidInputSec": 0,
    "WindowWidth": 1936,
    "WindowHeight": 1056,
    "MonitorWidth": 1920,
    "MonitorHeight": 1080,
    "MonitorFlags": 0,
    "WindowFlags": 18,
    "InteractiveTimeoutPeriodMS": 60000,
    "AggregationPeriodMS": 120000,
    "BitPeriodMS": 2000,
    "AggregationFlags": 49,
    "TotalUserOrDisplayActiveDurationMS": 12485,
    "SummaryRound": 0,
    "EventSequence": 16
}

**name: Win32kTraceLogging.AppInteractivitySummary**

This event was made when the Google chrome was running and the process name and the timestamp are verified at that time. A list of processes that were running at a particular time is able to be obtained, even though it is difficult for an application to specify the start and end times.

Figure 4. Examples of the telemetry data for property 1 (application life cycle: installation) and 3 (Process execution history).

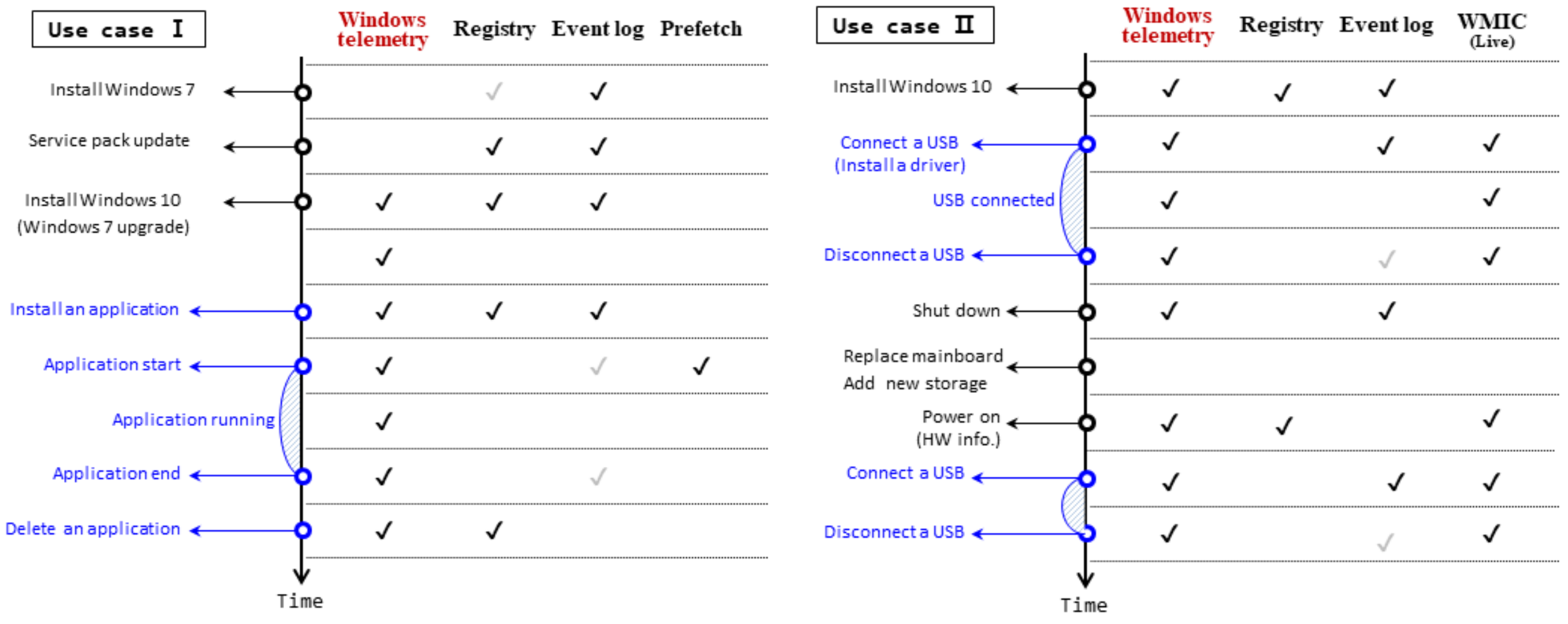

Figure 5. Comparison of the information that artifacts contains when events occurred: Windows telemetry, Registry, Event Log, Prefetch, WMIC (in case of live forensics).

## 4. Discussion

Windows telemetry RBS files are worth analyzing because telemetry data can be useful for understanding a target system. However, this is a new Windows artifact, so there are few previous digital forensic cases to refer to. Therefore, we present two situations as examples and discuss how the RBS files can be used in a digital forensic investigation as a fig. 5. This figure compares the history of the four artifacts when Windows is installed and updated, when system error occurs, when the application is installed, when process runs, and USB connection events occurs.

Windows artifacts (e.g. Registry, Prefetch, Event Log) have traces only when certain events occur due to user actions or system reservations but the RBS files store periodically and repeatedly regardless of the specific behavior. This differentiation enables telemetry data to give reliability to the results of analyzing other artifacts and additionally identifies traces that could not be identified prior to the traditional Windows forensics.

### 4.1 Case-study

**Incident response.** An incident response allows an organization to understand a data breach and prepare for future security incidents. In an incident response, collecting records of the event is the most fundamental requirement, and the system monitor (Sysmon) [21] and Google Rapid Response (GRR) [22] are widely used for this purpose. They provide detailed information about the incident, so an investigator can identify malicious or anomalous activity and understand how attackers or malware operated on the target system. The reason why RBS files are valuable is that they are a product of Windows 10 default settings and will thus be present even if Sysmon or GRR are not, both of which require an additional download. Of the properties of telemetry data described above, process execution history (Property 3) is probably the most useful. However, there is a possibility that an event can be missed. If the process is executed in a short time, it’s not possible to catch this events because the service does not collect the telemetry data in real time. When the MBR is damaged due to a cyberattack including a boot sector virus, it’s also possible to recover a system which can’t boot up using Property 5. In addition, the periodic configuration values of the other telemetry data allow to track changes in settings for an incident response.

**IT audit and Anti-forensics.** As a response to cyberattack such as intellectual property theft via unauthorized access, an IT audit can facilitate an understanding of the operating environment and the corresponding controls. During this type of investigation, it may be found that anti-forensic techniques such as a data hiding or artifact wiping have been employed. It is thus important to gain as many records as possible. If an auditor can acquire the same information from different artifacts, the reliability of the data improves; even if anti-forensic activity has been employed on one artifact, the missing information can be found in other artifacts. The RBS files provide a wealth of useful information in this regard. For example, over one year’s worth of information on the Windows update history and records of applications installed and uninstalled on the target system (Property 1) is available. If data leakage has occurred, tracing the connection of USB devices (Property 2) can also be used to identify the leak. If the suspect communicated with outsiders through a messenger service, the chat log can be decrypted using a hardware device identifier (Property 2). Furthermore, the timestamps in the telemetry data, in conjunction with the Event Log, can be used to track the running time of the target system, which can then act as a countermeasure against anti-forensic activity such as the modification of metadata.

### 4.2 Consideration and Future works

With a rise of IoT (Internet of Things) devices, telemetry is used throughout the industry and a related project [23] is underway for an effective observability. Diagnostics in Window 10 has also being updated constantly and the resulting the structure of the RBS file varied slightly and increased the types of the event depending on the version of the Windows Telemetry. Thus it needs a systematic support to manage a list of the telemetry data as a database to use telemetry data in digital forensic investigation, just like classifying Event Log by event ID. Only then will investigators be able to use Window telemetry as an artifact and to filter many events by the 'name' value. As shown in Table 1, the RBS file size is not very large because it is stored in a compressed data using the JSON file format, but it have more events than expected. Compared to the fact that NTFS filesystem logs are kept for less than a week on average experimentally, telemetry data has quite long trace for a target system or a disk dump image file because it has over the three months of process execution history and about one year of other properties. In this regard, an experiment is required for the cycle of an event depending on the key value of 'epoch'.

Additionally, we analyzed the RBS files but some of the data needs to be keep studying. In case of base64 encoded fields, reverse engineering analysis is required and then we will be able to determine whether personally identifiable information (PII) is included in order to address privacy concerns. Since PII have not found so far and Microsoft stated privacy principles with no personal content [24]. In fact, we've monitored network traffic that telemetry data is sent to Microsoft using TLS (Transport Layer Security) protocol that provides secure communications.

## 5. Conclusion

Microsoft has run a Windows telemetry service since the launch of Windows 10, collecting system data to enhance user experience. The traces collected as part of Windows telemetry is stored in the RBS files. We analyzed those data structure, and explored at least the valuable five properties, which the telemetry data contained history about the operating system and application installation, hardware device, process execution, external storage, and the boot sector. We also discussed how these can be used in digital forensic investigations. To date, Window telemetry RBS files have not been employed as artifacts in traditional Windows forensics. However, they contain a mix of old and new information and are thus worth analyzing.